\documentclass[twocolumn,showpacs,pra,amsmath,amssymb,eqsecnum]{revtex4}

\usepackage[T1]{fontenc}
\usepackage[latin9]{inputenc}

\usepackage{esint}
\usepackage{epsfig}
\usepackage{amsmath}
\usepackage{latexsym}

\usepackage{amssymb,amsmath}

\usepackage{graphicx}
\usepackage{dcolumn}
\usepackage{bm }
\usepackage{epstopdf}
\begin{document}

\def\be{\begin{equation}}
\def\ee{\end{equation}}

\title{Generating an electromagnetic multipole by oscillating currents}

\author{Asaf Farhi}
 \email{asaffarhi@post.tau.ac.il}
\author{David J. Bergman}
\email{bergman@post.tau.ac.il}

\affiliation{
Raymond and Beverly Sackler School of Physics and Astronomy,
Faculty of Exact Sciences,
Tel Aviv University, IL-69978 Tel Aviv, Israel
}

\date{\today}

\begin{abstract}

Based on the relation between a plane phased array and plane waves we show that a spherical current layer or a current sphere proportional
to a multipole electric field and situated in a uniform medium generates the same multipole field in all space. We calculate TE and TM multipoles inside and outside the spherical layer. The $l=1$ TM multipoles are localized at the origin with a focal spot with full width at half maximum of $0.4\lambda$ in the lateral axes and $0.58\lambda$ in the vertical axis. The multipole fields near the origin are prescriptions for the current distributions required to generate those multipole fields. A spherical layer can couple to a multipole source since the oscillation of the electrons in the layer due to the multipole field generates the multipole field in all space, which in turn can drive the multipole currents. Exciting a multipole in a polarizable sphere or spherical layer can couple it to another polarizable sphere or spherical layer.
\end{abstract}



\maketitle




\section{INTRODUCTION}
Scattering eigenstates of Maxwell's equations for a two-constituent composite
medium are fields that can exist in the system without an external 
source for a given permittivity value (eigenvalue) of one of the constituents, which will be called the inclusion.
When the outgoing waves are propagating the system radiates and that permittivity should have
gain. When the outgoing waves are evanescent and the system does
not radiate, that permittivity value is real.

By defining $\theta_{1}$ to be 1 inside the inclusion volume and zero otherwise, one can arrive from Maxwell's equations at the following equation for ${\bf E}({\bf r})$ in Gaussian units \cite{bergman1980theory}:
\begin{eqnarray}
\label{uk2}
-\nabla\times(\nabla\times{\bf E})+k_2^2{\bf E}=uk_2^2\theta_1{\bf E}-\frac{4\pi i\omega}{c^{2}}\mathbf{J}, \label{EdiffEq}\\
u \equiv 1-\frac{\epsilon_1}{\epsilon_2},\;\;\;\;k_2^2\equiv\epsilon_2\frac{\omega^2}{c^2},\nonumber
\end{eqnarray}
where $\epsilon_1,k_1$ are the permittivity and the wavevector of the inclusion, respectively, and $\epsilon_2, k_2$ are the permittivity and wavevector of the host medium. The eigenstates satisfy the following equation 
\begin{eqnarray}
\label{eigenstates}
-\nabla\times(\nabla\times{\bf E}_n)+k_2^2{\bf E}_n=u_nk_2^2\theta_1{\bf E}_n, \label{EdiffEq2}\\
u_n \equiv 1-\frac{\epsilon_{1n}}{\epsilon_2},\;\;\;\;\nonumber
\end{eqnarray}
where $\mathbf{E}_{n}$ has to satisfy the electric field continuity conditions at the interface characterized by $\theta_1.$

Since $\theta_{1}\mathbf{E}_{n}$ in Eq.\ (\ref{eigenstates}) plays the role
of current sources, an external current proportional to an eigenstate
in the inclusion volume situated in a homogeneous medium, denoted by $\mathbf{J}_{\textrm{ext}},$
will generate the two-constituent eigenstate, namely 

\begin{equation}
-\nabla\times(\nabla\times{\bf E}_n)+k_2^2{\bf E}_n= u_nk_2^2\theta_1{\bf J}_{\mathrm{ext}}. \label{EdiffEq3}
\end{equation}
In this equation the inclusion geometry is not specified
and this statement is therefore applicable to any inclusion geometry
and in particular for a flat slab, a cylinder,
and a sphere \cite{bergman1980theory,farhi2016electromagnetic,bergman2007electromagnetic}.

The electromagnetic eigenstates for a flat slab $-a<z<a$ in a host medium are of
the form \cite{farhi2016electromagnetic}
\[
\mathbf{E}_{\mathbf{k}\,\mathrm{TE}}^{+}=e^{i\mathbf{k}\cdot\boldsymbol{\rho}}\left\{ \begin{array}{cc}
\mathbf{e}_{\perp}A_{\perp}^{+}e^{-ik_{2z}z} & z>a\\
\mathbf{e}_{\perp}B_{\perp}^{+}\cos\left(k_{1z}^{+}z\right) & -a<z<a\\
\mathbf{e}_{\perp}A_{\perp}^{+}e^{ik_{2z}z} & z<-a
\end{array}\right.,
\]
where an even (`+') TE eigenstate is presented,  $\hat{\mathbf{x}},\hat{\mathbf{y}}$ and $\hat{\mathbf{z}}$ are parallel and perpendicular to the slab, respectively,
\mbox{$k_{1z}^+\equiv \sqrt{(k_1^+)^2-{\bf k}^2},$}\mbox{$\,\,\,k_{2z}\equiv \sqrt{(k_2)^2-{\bf k}^2},$}
\mbox{$k_1^+\equiv\sqrt{\epsilon^+_{1k}}\omega/c,$} \mbox{$k_2\equiv\sqrt{\epsilon_2}\omega/c,$} $\mathbf{e}_{\perp}=\mathbf{e}_{\mathbf{k}}\times\mathbf{e}_{z},$ \mbox{$\mathbf{e}_{\mathbf{k}}\equiv{\bf k}/|{\bf k}|,$} {\bf k} is a real 2D wave vector in a direction parallel to the slab,  $\boldsymbol{\rho}$ is a 2D position
vector in that plane, and $\epsilon^+_{1k}$ is the $k$ dependent slab permittivity eigenvalue.
Thus, $\mathbf{J}_{\textrm{ext}}\propto\theta_{1}e^{i\mathbf{k}\cdot\boldsymbol{\rho}}\mathbf{e}_{\perp}B_{\perp}^{+}\cos\left(k_{1z}^{+}z\right)$
in a homogeneous medium will generate this eigenstate in all space. 

Phased arrays are arrays of antennas with predetermined phases which generate a desired electromagnetic wave, usually a plane
wave propagating in a given direction. Optical antenna arrays are based
on the polarization of a material due to an applied electric field
and act as current sources at optical frequencies \cite{novotny2011antennas}. 

To understand how a plane phased array can generate a plane wave
we can operate with the free-space Green's tensor on a continuous
current sheet $\mathbf{J}_{\textrm{ext}}=\mathbf{e}_{\perp}e^{-i\mathbf{k}\cdot\boldsymbol{\rho}}\delta\left(z\right)$ to obtain
\begin{align}
&\overleftrightarrow{G}\ast\left[\mathbf{e}_{\perp}e^{-i\mathbf{k}\cdot\boldsymbol{\rho}}\delta\left(z\right)\right]\propto \nonumber \\
&\mathbf{e}_{\perp}\left[\theta(z)e^{-i\mathbf{k}\cdot\boldsymbol{\rho}+ik_{2z}z}+\theta(-z)e^{-i\mathbf{k}\cdot\boldsymbol{\rho}-ik_{2z}z}\right],
\end{align}
where Green's tensor is defined by the following equation
\[
-\nabla\times\nabla\times\overleftrightarrow{G}+k_{2}^{2}\overleftrightarrow{G}=k_{2}^{2}I\delta^3\left(\mathbf{r}-\mathbf{r}'\right)
\]
and is an outgoing wave at large distances.
Here $\theta(z)$ is the usual step function, {\bf k} is a 2D wave vector parallel to the plane, and $\boldsymbol{\rho}$ is a 2D position
vector in that plane. Thus, a current sheet which is proportional to a plane wave modulated by $e^{-i\mathbf{k}\cdot\boldsymbol{\rho}},$ 
generates outgoing plane waves away from the current sheet with the
same 2D phase modulation. It will be therefore interesting to check
if the same applies for sphere and cylinder geometries. Light usually comprises a variety of multpoles \cite{orlov2012analytical} and our goal is to generate only a given multipole.

Vector spherical harmonics with the electric field continuity conditions
are the eigenstates of a spherical inclusion \cite{bergman1980theory}. Standard vector spherical harmonics are the eigenstates of a uniform medium similarly to plane waves. The analogue of a planar phased array is therefore a spherical current layer modulated by a vector spherical harmonic. 
Light can be localized in free space by focusing a laser beam using a circular lens. The size of the focal spot is related to the imaging resolution since the scattered light can be collected from the focal spot, enabling to resolve features with the size of the focal spot. A laser beam with a uniform distribution passing through a circular lens has a focal spot with a full width at half maximum (FWHM) of $1.03\lambda f/D$ in the lateral axes, where $f$ is the focal length and $D$ is the lens diameter (commercially available lenses have $f/D\gtrsim 1$). Setups of molecules coupled to optical cavities have attracted much attention recently. In the weak coupling regime the emission of the molecule can be enhanced and in the strong coupling regime the system has hybrid eigenstates \cite{chikkaraddy2016single,thompson2013coupling}. The electric field  of a molecule (modeled by an oscillating dipole) in a dielectric sphere  has been calculated in the electrodynamic regime \cite{van1937xiii,chew1976model}. Radiation emission rates for such a setup which also accounts for electrostatic spherical and spheroidal cavity effects have also been calculated \cite{gersten1991radiative}. Here we show that a spherical current layer or a sphere modulated by a vector spherical harmonic can generate the corresponding spherical harmonic. In addition, a spherical current layer proportional to the $l=1$ TM mode can focus light at the origin. Such currents at optical frequencies can be effectively generated by optical antennas. Moreover, we show that a polarizable spherical layer can couple to a multipole source. In Sec.\ \ref{sec:theory} we present the theory. In Sec.\ \ref{sec:multipole_fields} we calculate multipole fields and present results near the origin and in the far field. In Sec.\ \ref{sec:discussion} we discuss our results.

\section{Theory}
\label{sec:theory}
The electric field can be expanded in the multipoles as follows
\begin{eqnarray}
\mathbf{E}=\sum_{l,m}\left[i/k\cdot a_{E}\left(l,m\right)\nabla\times f_{l}\left(kr\right)\mathbf{X}_{lm}\right.\nonumber\\
 +\left.a_{M}\left(l,m\right)g_{l}\left(kr\right)\mathbf{X}_{lm}\right],
 \end{eqnarray}
 where $a_{E},\,a_{M}$ are the multipole strengths, $f_{l}\left(kr\right),\,g_{l}\left(kr\right)$ are linear combinations of radial Hankel functions $h_{l}^{\left(1\right)},h_{l}^{\left(2\right)}$, and $g_{l}\left(kr\right)\mathbf{X}_{lm},\nabla\times f_{l}\left(kr\right)\mathbf{X}_{lm}$ are the normalized forms of the TE and TM electric field multipoles $\mathbf{E}_{lm}^{\left(M\right)}$ and $\mathbf{E}_{lm}^{\left(E\right)}$ (the vector spherical harmonics) given by 
\begin{align}
&\mathbf{E}_{lm}^{\left(M\right)}=g_l\left(kr\right)\mathbf{L}Y_{lm}\left(\theta,\phi\right), \mathbf{\,\,\, L}\equiv\frac{1}{i}\left(\mathbf{r}\times\nabla\right),\nonumber \\
&\mathbf{E}_{lm}^{\left(E\right)}=\frac{i}{k}\nabla\times\mathbf{H}_{lm}^{\left(E\right)},\,\,\,\,  \mathbf{H}_{lm}^{\left(E\right)}\equiv f_{l}\left(kr\right)\mathbf{L}Y_{lm}\left(\theta,\phi\right),
\end{align}
where $Y_{lm}\left(\theta,\phi\right)$ are the scalar spherical harmonics.

The multipole strengths $\alpha_{M}\left(l,m\right),\alpha_{E}\left(l,m\right)$ can be written
as follows 
\begin{align}
\alpha_{M}\left(l,m\right)g_{l}\left(kr\right)=\frac{k}{\sqrt{l\left(l+1\right)}}\int Y_{lm}^{*}\left(\mathbf{r}\cdot\mathbf{H}\right)d\Omega,\nonumber \\
\alpha_{E}\left(l,m\right)f_{l}\left(kr\right)=-\frac{k}{\sqrt{l\left(l+1\right)}}\int Y_{lm}^{*}\left(\mathbf{r}\cdot\mathbf{E}\right)d\Omega.
\end{align}
Denoting the intrinsic magnetization (caused by circular currents)
by $\mathbf{M}\left(\mathbf{x}\right)$ the following scalar wave equation for $\mathbf{r}\cdot\mathbf{H},\,\,\,\mathbf{r}\cdot\mathbf{E'}$, where $\mathbf{E'}$ is the divergence-free field defined by $\mathbf{E'}=\mathbf{E}+\frac{i}{\omega\epsilon_0}\mathbf{J}$ (note that $\nabla\cdot\mathbf{E}=\rho/\epsilon_{0},\,\,\,\,\nabla\cdot\mathbf{J}=i\omega\rho$), can be written as 
\begin{align}
&\left(\nabla^{2}+k^{2}\right)\left(\mathbf{r}\cdot\mathbf{H}\right)=-i\mathbf{L}\cdot\left(\mathbf{J}+\nabla\times\mathbf{M}\right),\nonumber \\
& \left(\nabla^{2}+k^{2}\right)\left(\mathbf{r}\cdot\mathbf{E}'\right)=k\mathbf{L}\cdot\left(\mathbf{M}+\frac{1}{k^{2}}\nabla\times\mathbf{J}\right)
\end{align}
from which one can write \cite{jackson1975electrodynamics}
\begin{align}
&\mathbf{r}\cdot\mathbf{H}=\frac{i}{4\pi}\int\frac{e^{ik\left|\mathbf{x}-\mathbf{x}'\right|}}{\left|\mathbf{x}-\mathbf{x}'\right|}\mathbf{L}'\cdot\left[\mathbf{J}\left(\mathbf{x}'\right)+\nabla\times\mathbf{M}\left(\mathbf{x}'\right)\right]d^{3}x',\nonumber \\
&\mathbf{r}\cdot\mathbf{E}'=\nonumber \\
&-\frac{k}{4\pi}\int\frac{e^{ik\left|\mathbf{x}-\mathbf{x}'\right|}}{\left|\mathbf{x}-\mathbf{x}'\right|}\mathbf{L}'\cdot\left(\mathbf{M}\left(\mathbf{x}'\right)+\frac{1}{k^{2}}\nabla\times\mathbf{J}\left(\mathbf{x}'\right)\right)d^{3}x',
\end{align}
where in the region outside the sources $\mathbf{E}'$ reduces to
$\mathbf{E}.$

Now $\alpha_{M}\left(l,m\right),\alpha_{E}\left(l,m\right)$ read
\begin{widetext}
\begin{align}
\alpha_{M}\left(l,m\right)g_{l}\left(kr\right)&=\frac{i}{4\pi}\frac{k}{\sqrt{l\left(l+1\right)}}\int Y_{lm}^{*}\int\frac{e^{ik\left|\mathbf{x}-\mathbf{x}'\right|}}{\left|\mathbf{x}-\mathbf{x}'\right|}\mathbf{L}'\cdot\left[\mathbf{J}\left(\mathbf{x}'\right)+\nabla\times\mathbf{M}\left(\mathbf{x}'\right)\right]d^{3}x'd\Omega,\nonumber\\
\alpha_{E}\left(l,m\right)f_{l}\left(kr\right)&=\frac{1}{4\pi}\frac{k^{2}}{\sqrt{l\left(l+1\right)}}\int Y_{lm}^{*}\int\frac{e^{ik\left|\mathbf{x}-\mathbf{x}'\right|}}{\left|\mathbf{x}-\mathbf{x}'\right|}\mathbf{L}'\cdot\left[\mathbf{M}\left(\mathbf{x}'\right)+\frac{1}{k^2}\nabla\times\mathbf{J}\left(\mathbf{x}'\right)\right]d^{3}x'd\Omega.
\label{eq:alphamg}
 \end{align}
We substitute $\frac{e^{ik\left|\mathbf{x}-\mathbf{x}'\right|}}{4\pi\left|\mathbf{x}-\mathbf{x}'\right|}\equiv \sum_{l,m}\mathcal{G}_{l}\left(r,r'\right)Y_{l,m}^{*}\left(\theta',\phi'\right)Y_{l,m}\left(\theta,\phi\right),$
where $\mathcal{G}_{l}\left(r,r'\right)=ikj_{l}\left(kr_{<}\right)h_{l}^{\left(1\right)}\left(kr_{>}\right),\,r_{<}\left(r_{>}\right)$ is the smaller (larger) of $r$ and $r'.$ Note that $g_l(kr)=f_l(kr)=h_{l}^{\left(1\right)}(kr)$ outside the current layer $(r>r'),$ and $g_l(kr)=f_l(kr)=j_l(kr)$ inside $(r<r')$ so one of the functions in $\mathcal{G}_{l}\left(r,r'\right)$  cancels out with $g_l(kr)$ or $f_l(kr)$ on the lhs of Eq.\ (\ref{eq:alphamg}). Consequently  $\alpha_{M}\left(l,m\right),\alpha_{E}\left(l,m\right)$ inside and outside take the form 
\begin{align}
\alpha^{\mathrm{inside}/\mathrm{outside}}_{M}\left(l,m\right)&=-\frac{k^2}{\sqrt{l\left(l+1\right)}}\int \left\{ \begin{array}{c}
h_{l}^{\left(1\right)}\left(kr\right)\\
j_{l}\left(kr\right)
\end{array}\right\} 
  Y_{lm}^{*}\left(\theta,\phi\right)\mathbf{L}\cdot\left[\mathbf{J}\left(\mathbf{x}\right)+\nabla\times\mathbf{M}\left(\mathbf{x}\right)\right]d^{3}x,\nonumber \\
\alpha^{\mathrm{inside}/\mathrm{outside}}_{E}\left(l,m\right)&=\frac{ik^{3}}{\sqrt{l\left(l+1\right)}}\int \left\{ \begin{array}{c}
h_{l}^{\left(1\right)}\left(kr\right)\\
j_{l}\left(kr\right)
\end{array}\right\} 
  Y_{lm}^{*}\left(\theta,\phi\right)\mathbf{L}\cdot\left[\mathbf{M}\left(\mathbf{x}\right)+\frac{1}{k^2}\nabla\times\mathbf{J}\left(\mathbf{x}\right)\right]d^{3}x,
\label{eq:strengths}
\end{align}
\end{widetext}
where ``inside'' and ``outside'' correspond to $h_{l}^{\left(1\right)}\left(kr\right)$ and $j_{l}\left(kr\right),$ respectively.
We consider a spherical current layer situated in free space with an inner radius $r_{1}$ and a thickness $d$
where for the case of a surface current $d\ll r_{1}.$ The current is proportional to a TE multipole in its $\theta,\phi$ dependence in the spherical layer $\mathbf{J}=J_0 \theta_1 \left(r\right)  w_{l'}\left(kr\right) \mathbf{L}Y_{l'm'}\left(\theta,\phi\right)$ where $\theta_1$ is equal to 1 when $r$ is inside the current layer and is equal to zero elsewhere.
 However, the function $w_{l'}\left(kr\right)$ in general is not a spherical Bessel function.
  We assume $\mathbf{M}\left(\mathbf{x}\right)=0$ from here on and obtain 
\begin{align}
\alpha^{\mathrm{inside}/\mathrm{outside}}_{M}&\left(lm,l'm'\right)= -k^{2}\sqrt{l\left(l+1\right)}J_0 \delta_{ll'}\delta_{mm'} \times \nonumber\\
 &\intop_{r_{1}}^{r_{1}+d}\left\{ \begin{array}{c}
h_{l}^{\left(1\right)}\left(kr\right)\\
j_{l}\left(kr\right)
\end{array}\right\} w_{l'}\left(kr\right)r^{2}dr.
\end{align}
It can be seen that such a surface current source generates only the same TE multipole.

From here on we will consider $r$ that is inside and the calculations for $r$ outside readily follow, namely $h_{l}^{\left(1\right)}(kr)$ should be replaced by $j_{l}\left(kr\right)$ in the expressions for the multipole strengths.
To prove that TM multipoles are not generated by TE surface currents we perform several mathematical operations and arrive from Eq. (\ref{eq:strengths}) at
\begin{widetext}
\begin{align}
&a^{\mathrm{inside}}_{E}\left(lm,l'm'\right)=\frac{k^{2}}{i\sqrt{l\left(l+1\right)}}\int Y_{lm}^{*}\left[c\rho\frac{\partial}{\partial r}\left[rh_{l}^{\left(1\right)}\left(kr\right)\right]+ik\left(\mathbf{r}\cdot\mathbf{J}\right)h_{l}^{\left(1\right)}\left(kr\right)\right]drd\Omega \nonumber\\
&=\frac{k^{2}}{i\sqrt{l\left(l+1\right)}}\frac{c}{i\omega}\left[\int\intop_{r_{1}-\frac{dr}{2}}^{r_{1}+\frac{dr}{2}}\left\{ \nabla\cdot\left[Y_{lm}^{*}\frac{\partial}{\partial r}\left[rh_{l}^{\left(1\right)}\left(kr\right)\right]\mathbf{J}\right]-\mathbf{J}\cdot\nabla\left[Y_{lm}^{*}\frac{\partial}{\partial r}\left[rh_{l}^{\left(1\right)}\left(kr\right)\right]\right]\right\} r^{2}drd\Omega\right.\nonumber\\
&+\left.\int\intop_{r_{1}+d-\frac{dr}{2}}^{r_{1}+d+\frac{dr}{2}}\left\{ \nabla\cdot\left[Y_{lm}^{*}\frac{\partial}{\partial r}\left[rh_{l}^{\left(1\right)}\left(kr\right)\right]\mathbf{J}\right]-\mathbf{J}\cdot\nabla\left[Y_{lm}^{*}\frac{\partial}{\partial r}\left[rh_{l}^{\left(1\right)}\left(kr\right)\right]\right]\right\} r^{2}drd\Omega\right]=-\frac{ck^{2}}{\omega\sqrt{l\left(l+1\right)}}\times \nonumber\\
&\int Y_{lm}^{*}\left[\mathbf{J}_{\textrm{above}\,\textrm{upper}}\cdot\hat{\mathbf{r}}\left[\frac{\partial}{\partial r}\left(rh_{l}^{\left(1\right)}\left(kr\right)\right)\right]_{r_{1}+d}\left(r_{1}+d\right)^{2}-\mathbf{J}_{\textrm{below}\,\textrm{upper}}\cdot\hat{\mathbf{r}}\left[\frac{\partial}{\partial r}\left(rh_{l}^{\left(1\right)}\left(kr\right)\right)\right]_{r_{1}+d}\left(r_{1}+d\right)^{2}\right.\nonumber\\
&\left.\mathbf{J}_{\textrm{above}\,\textrm{lower}}\cdot\hat{\mathbf{r}}\left[\frac{\partial}{\partial r}\left(rh_{l}^{\left(1\right)}\left(kr\right)\right)\right]_{r_{1}}r_{1}^{2}-\mathbf{J}_{\textrm{below}\,\textrm{lower}}\cdot\hat{\mathbf{r}}\left[\frac{\partial}{\partial r}\left(rh_{l}^{\left(1\right)}\left(kr\right)\right)\right]_{r_{1}}r_{1}^{2}\right]d\Omega=0,
\end{align}
\end{widetext}
where $\hat{\mathbf{r}}=\mathbf{r}/|\mathbf{r}|,\,\,\,\mathbf{r}\cdot\mathbf{J\propto\mathbf{r}\cdot}\mathbf{E}_{l'm'}^{\left(M\right)}=0,$ $\rho=\frac {\nabla\cdot\mathbf{J}}{i\omega},$ inside the volume $\nabla\cdot\mathbf{J}=0$ since $\mathbf{J}\propto  \theta_1\left(r\right)\mathbf{E}_{l'm'}^{\left(M\right)},$ $\psi\nabla\cdot\mathbf{J}=\nabla\cdot\left(\psi\mathbf{J}\right)-\mathbf{J}\cdot\nabla\psi,\,\,\psi\equiv Y_{lm}^{*}\frac{\partial}{\partial r}\left[rh_{l}^{\left(1\right)}\left(kr\right)\right],$ the volume integrals over the interfaces transform to surface integrals, `upper' and `lower' denote the upper and lower interface respectively, `above' and `below' stand for above and below an interface, and $\mathbf{J}_{\textrm{above}\,\textrm{upper}}=\mathbf{J}_{\textrm{below}\,\textrm{lower}}=0.$ Also, the contribution of the $\mathbf{J}\cdot\nabla\left[Y_{lm}^{*}\frac{\partial}{\partial r}\left[rh_{l}^{\left(1\right)}\left(kr\right)\right]\right]$ terms in the volume integration over the interfaces is negligible since the interface thickness is infinitesimal and the function is finite there.
  
We define $\tilde{J}_{\theta}=\frac{J_{\theta}}{w_{l'}\left(kr\right)},\tilde{J}_{\phi}=\frac{J_{\phi}}{w_{l'}\left(kr\right)}$
 and since $J_{r}=0$ we express $\nabla\cdot\mathbf{J}$ inside the volume as follows
\begin{align}
\nabla\cdot\mathbf{J}&=\frac{1}{r\sin\theta}\left[\frac{\partial}{\partial\theta}\left(\sin\theta J_{\theta}\right)+\frac{\partial J_{\phi}}{\partial\phi}\right]\nonumber \\
&=\frac{w_{l'}\left(kr\right)}{r\sin\theta}\left[\frac{\partial}{\partial\theta}\left(\sin\theta\tilde{J}_{\theta}\right)+\frac{\partial\tilde{J}_{\phi}}{\partial\phi}\right]=0.
\end{align}
We get that since $\frac{\partial}{\partial\theta}\left(\sin\theta\tilde{J}_{\theta}\right)+\frac{\partial\tilde{J}_{\phi}}{\partial\phi}=0,$ the current satisfies $\nabla\cdot\mathbf{J}=0$ independently of the form of $w_{l'}\left(kr\right).$ Here, too, $w_{l'}\left(kr\right)$ is not required to be a spherical Bessel function.

Similarly, substituting a spherical layer source proportional to
a TM multipole
\begin{align}
\mathbf{J}&=\frac{i}{k}J_0\theta_{1}\left(r\right)\nabla\times f_{l'}\left(kr\right)\mathbf{L}Y_{l'm'}\left(\theta,\phi\right) \nonumber\\
&\propto\theta_{1}\left(r\right)\mathbf{E}_{l'm'}^{\left(E\right)}\propto\theta_{1}\left(r\right)\frac{i}{k}\nabla\times\mathbf{H}^{\left(E\right)}_{l'm'},
\end{align}
we obtain
\begin{widetext}
\begin{align}
a^{\mathrm{inside}}_{E}\left(lm,l'm'\right)&=\frac{k^{2}}{i\sqrt{l\left(l+1\right)}}\int Y_{lm}^{*}\left[c\rho\frac{\partial}{\partial r}\left[rh_{l}^{\left(1\right)}\left(kr\right)\right]+iJ_0l'\left(l'+1\right)f_{l'}\left(kr\right)Y_{l'm'}h_{l}^{\left(1\right)}\left(kr\right)d^{3}x\right] \nonumber \\
&=\frac{k^{2}}{i\sqrt{l\left(l+1\right)}}\int Y_{lm}^{*}\left\{ -\frac{c}{i\omega}\mathbf{J}\left(r_{1}+d,\theta,\phi\right)\cdot\hat{\mathbf{r}}\left[\frac{\partial}{\partial r}\left(rh_{l}^{\left(1\right)}\left(kr\right)\right)\right]_{r_{1}+d}\left(r_{1}+d\right)^{2}\right.\nonumber\\
&+\frac{c}{i\omega}\mathbf{J}\left(r_{1},\theta,\phi\right)\cdot\hat{\mathbf{r}}\left[\frac{\partial}{\partial r}\left(rh_{l}^{\left(1\right)}\left(kr\right)\right)\right]_{r_{1}}r_{1}^{2}+\left.iJ_0l'\left(l'+1\right)f_{l'}\left(kr\right)Y_{l'm'}h_{l}^{\left(1\right)}\left(kr\right)r^2 dr\right\} d\Omega\nonumber\\
&=J_{0}k\sqrt{l\left(l+1\right)}\delta_{ll'}\delta_{mm'}\left\{\frac{c}{\omega} \frac{f_{l'}\left(k\left(r_{1}+d\right)\right)}{\left(r_{1}+d\right)}\left[\frac{\partial}{\partial r}\left(rh_{l}^{\left(1\right)}\left(kr\right)\right)\right]_{r_{1}+d}\left(r_{1}+d\right)^{2}\right.\nonumber \\
&\left.-\frac{c}{\omega}\frac{f_{l'}\left(kr_{1}\right)}{ r_{1}}\left[\frac{\partial}{\partial r}\left(rh_{l}^{\left(1\right)}\left(kr\right)\right)\right]_{r_{1}}r_{1}^{2}+k\intop_{r_{1}}^{r_{1}+d} f_{l'}\left(kr\right)h_{l}^{\left(1\right)}\left(kr\right)r^{2}dr\right\},
\end{align}
\end{widetext}
where we have used $\mathbf{r}\cdot\mathbf{J}=J_0\frac{l'\left(l'+1\right)}{k}f_{l'}\left(kr\right)Y_{l'm'}\propto\mathbf{r}\cdot\mathbf{E}_{l'm'}^{\left(E\right)}$ for $\mathbf{r}\cdot\mathbf{J},\,\,\,\mathbf{J}\left(r_{1}+d,\theta,\phi\right)\cdot\hat{\mathbf{r}},\,\mathrm{and}\,\,\mathbf{J}\left(r_{1},\theta,\phi\right)\cdot\hat{\mathbf{r}}$, the orthogonality property of the $Y_{lm}$s, and $\nabla\cdot\mathbf{J}=0$ inside the current volume since $\mathbf{J}\propto\mathbf{E}_{l'm'}^{\left(E\right)}.$

Since for every vector $\nabla\cdot\nabla\times\mathbf{A}=0,$ it is easy to see from the definition of the TM multipole that we can replace $f_{l'}\left(kr\right)$ with any $r$ dependent function and the derivation above still holds.


To prove that a TM surface current does not generate TE multipoles we write
\begin{widetext}
\begin{equation}
\alpha^{\mathrm{inside}}_{M}\left(lm,l'm'\right)=\frac{-k^{2}}{\sqrt{l\left(l+1\right)}}\int h_{l}^{\left(1\right)}\left(kr\right)Y_{lm}^{*}\left(\theta,\phi\right)\mathbf{L}\cdot\frac{i}{k}J_0\theta_{1}\left(r\right)\nabla\times\left[f_{l'}\left(kr\right)\mathbf{L}Y_{l'm'}\left(\theta,\phi\right)\right]d^{3}x =0,
\label{TM_no_TE_mult}
\end{equation}
\end{widetext}
where we have used 
\begin{align}
&\,\,\,\,\,\,\,\,\,\,\,\,\,\,\,\,\,\,\,\,\,\,\,\,\,\,\mathbf{L}=\frac{1}{i}\left(\mathbf{r}\times\nabla\right), \nonumber\\
&\,\,\,\,\,\,\left(\mathbf{r}\times\nabla\right)\cdot\nabla\times\left[f_{l'}\left(kr\right)\mathbf{L}Y_{l'm'}\left(\theta,\phi\right)\right]  \nonumber\\
&=\mathbf{r}\cdot\nabla\times\left\{ \nabla\times\left[f_{l'}\left(kr\right)\mathbf{L}Y_{l'm'}\left(\theta,\phi\right)\right]\right\}   \nonumber\\
&=-\mathbf{r}\cdot\left[k^{2}f_{l'}\left(kr\right)\mathbf{L}Y_{l'm'}\left(\theta,\phi\right)\right]=0,  \nonumber
 \end{align}
since $\nabla\cdot\mathbf{E}_{l'm'}^{\left(E\right)}=0,\nabla^2\mathbf{E}_{l'm'}^{\left(E\right)}=-k^2\mathbf{E}_{l'm'}^{\left(E\right)},$ and since $\mathbf{L}$ does not operate on $r.$
 Here, too, a surface current source which is proportional
to a TM multipole generates only the same multipole. Clearly, the results for the strengths of the multipoles are also valid for a spherical current layer that is thick and for a sphere, where we are interested in  $\alpha^{\mathrm{outside}}_{M}\left(l,m\right),\alpha^{\mathrm{outside}}_{E}\left(l,m\right).$
\section{Multipole fields near the origin and in the far field}
\label{sec:multipole_fields}
\begin{figure}[t]
\includegraphics[width=8cm]{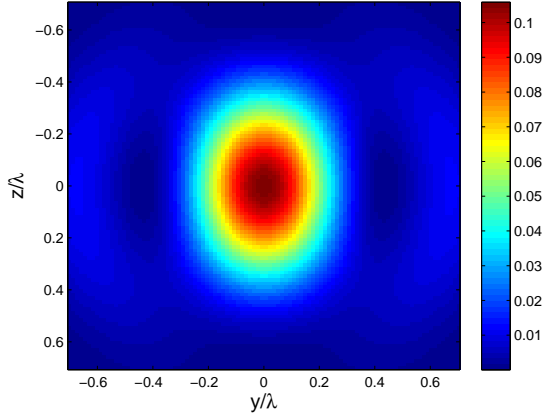}
\protect\caption{\label{fig:1} $\left|\mathbf{E}\right|^2$ of a TM $l=1,m=0$ multipole}
\end{figure}

\begin{figure}[h]
\includegraphics[width=8cm]{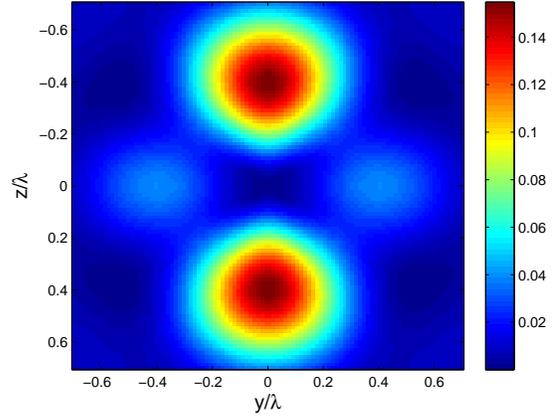}
\protect\caption{\label{fig:2}$\left|\mathbf{E}\right|^2$ of a TM $l=2,m=0$ multipole}
\end{figure}

We calculated the $l=1,m=0$ and the $l=2,m=0$ TM and TE multipole fields which can be generated by the corresponding spherical current layers.
We first calculated the $l=1,m=0$ and the $l=2,m=0$ TM multipole fields with $f_{l}\left(kr\right)\equiv j_{l}\left(kr\right)$
  using the following relations \cite{jackson1975electrodynamics}
\begin{align}
&\nabla\times\mathbf{L}=\mathbf{r}\nabla^{2}-\nabla, \,\,\,L_{x}=\frac{1}{2}\left(L_{+}+L_{-}\right),\nonumber\\ 
&L_{y}=\frac{1}{2i}\left(L_{+}-L_{-}\right),\,\,\,L_{z}=mY_{lm},\nonumber\\
&L_{+}Y_{l,m}=\sqrt{\left(l-m\right)\left(l+m+1\right)}Y_{l,m+1},\nonumber\\
&L_{-}Y_{l,m}=\sqrt{\left(l+m\right)\left(l-m+1\right)}Y_{l,m-1}.
\end{align}
\noindent
In Fig.\ \ref{fig:1} we present  $\left|\mathbf{E}\right|^2$ of the $l=1,m=0$ TM multipole as a function of $y$ and $z$ near the origin. The electric field of this multipole is concentrated at the origin and has a FWHM of $0.4\lambda$ in the $y$ axis (and in the $x$ axis) and $0.58\lambda$ in the $z$ axis. Note that while to first order $j_{l=1}\left(kr\right)\simeq (kr)/3,$ upon operating with $\nabla \times$ on $j_{l=1}\left(kr\right)\mathbf{L}Y_{lm}$ in the calculation of $\mathbf{E}_{l=1,m}^{\left(E\right)},$ the field does not vanish at the origin. In Fig.\ \ref{fig:2} we present $\left|\mathbf{E}\right|^2$ of the $l=2,m=0$ TM multipole. This multipole has two focal spots at $z=\pm0.4\lambda$ with focal width of approximately $0.4\lambda.$ 

\begin{figure}[h]
\includegraphics[width=8cm]{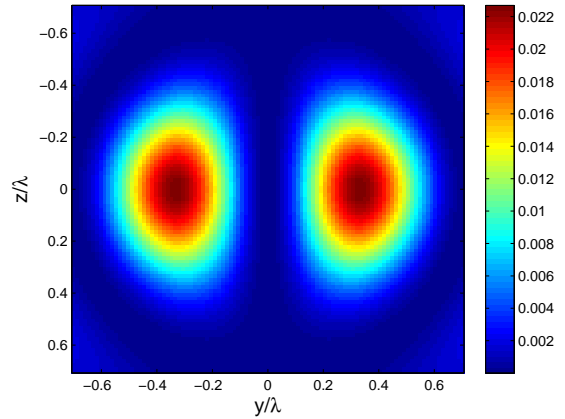}
\protect\caption{\label{fig:3} $\left|\mathbf{E}\right|^2$ of a TE $l=1,m=0$ multipole}
\end{figure}

\begin{figure}[h]
\includegraphics[width=8cm]{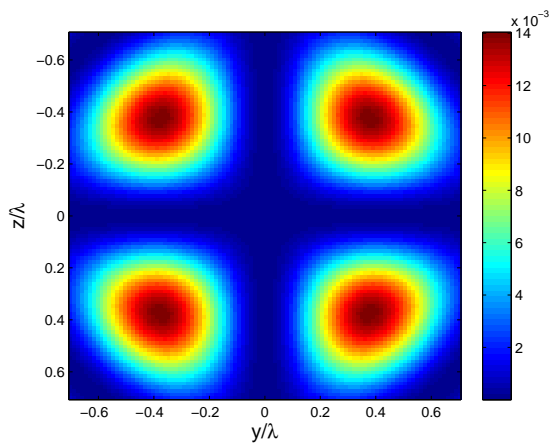}
\protect\caption{\label{fig:4} $\left|\mathbf{E}\right|^2$ of a TE $l=2,m=0$ multipole}
\end{figure}

We then calculated $\left|\mathbf{E}\right|^2$ of the $l=1,m=0$ and $l=2,m=0$ TE multipoles with $g_{l}\left(kr\right)\equiv j_{l}\left(kr\right)$ by using the relation $L^{2}=l\left(l+1\right)$ and obtained
\begin{align}
&\left|\mathbf{E}_{\mathrm{TE}\,1,0}\right|^{2}=\frac{3}{8\pi}\left(j_{1}\left(kr\right)\right)^{2}\sin^{2}\theta,\nonumber\\
&\left|\mathbf{E}_{\mathrm{TE}\,2,0}\right|^{2}=\frac{15}{8\pi}\left(j_{2}\left(kr\right)\right)^{2}\sin^{2}\theta\cos^{2}\theta.
 \end{align}
In Fig.\ \ref{fig:3} we present $\left|\mathbf{E}\right|^2$ of the $l=1,m=0$ TE multipole as a function of $y$ and $z$ near the origin. The intensity of this multipole is concentrated in a torus-like shape. In Fig.\ \ref{fig:4} we present $\left|\mathbf{E}\right|^2$ of the $l=2,m=0$ TE multipole. The intenisty of this multipole is concentrated in two torus-like shapes situated at $z=\pm0.4\lambda.$  

From the derivations in Sec.\ \ref{sec:theory} it is clear that placing currents that are modulated according to the $l=1,m=0$ TE and TM multipoles inside a sphere near the origin generates the electric field of these modes in space. Thus, electric fields whose intensities are presented in Figs.\ \ref{fig:1} and \ref{fig:3}, are prescriptions for the current distributions required to generate the $l=1,m=0$ TM and TE multipoles respectively. The $l=1,m=0$ TM multipole field in a very small sphere corresponds to an oscillating point dipole. In addition, the focal spot of this mode is similar to the current distribution of a $\lambda/2$ antenna, which has a dominant $l=1,m=0$ TM component in its radiation field pattern. The $l=1,m=0$ TE multipole field is similar to a current loop distribution. 

In addition, a multipole current source at the origin and a spherical layer centered at the origin can be coupled. The radiation emitted from the multipole source impinges on the spherical layer, which in turn is polarized and emits radiation both inside and outside that layer. The radiation inside impinges on the current source and can re-excite the same radiation pattern etc. The polarizable spherical layer is characterized by an $\epsilon$ value and can be either a dielectric or a metal. The polarization pattern in the spherical layer originates from the radiation field pattern of the current source and there is no need to design an optical antenna distribution inside the layer. Similarly if a polarizable spherical layer or a sphere is excited with a given multipole it will couple to another sphere or spherical layer centered at the origin for the reason mentioned above. 

In the radiation zone we can write $\mathbf{E}_{lm}^{\left(M\right)}=\mathbf{H}_{lm}^{\left(M\right)}\times\mathbf{n}.$
For a source which consists of a complete set of TE multipoles with the order $l$ all having the same strength $a_{M}\left(l\right),$ the time-averaged power radiated per solid angle is
\begin{equation}
\frac{dP}{d\Omega}=\frac{1}{2k^{2}}\left|a_{M}\left(l\right)\right|^{2}\left|\sum_{m}\left(-1\right)^{i+1}\mathbf{X}_{lm}\right|^{2},
\end{equation}
which becomes for incoherent sources \cite{jackson1975electrodynamics}
\begin{equation}
\frac{dP}{d\Omega}=\frac{1}{2k^{2}}\left|a_{M}\left(l\right)\right|^{2}\frac{2l+1}{4\pi}.
\end{equation}
Thus, generating such a set of incoherent TE multipoles will result in isotropic radiation, where the simplest configuration comprises the multipoles $\mathbf{X}_{10},\mathbf{X}_{1,-1},\mathbf{X}_{1,1}.$
 \\
 \\
\section{Discussion}
\label{sec:discussion}
We showed that a spherical layer or a sphere with currents that are proportional to an electromagnetic multipole field can generate the same multipole field. We calculated the first TE and TM multipole fields and presented results near the origin and in the far field. The intensity of the $l=1,m=0$ TM multipole peaks at the origin with FWHM of $0.4\lambda$ in the lateral axes and $0.58\lambda$ in the vertical axis and the intensity of the $l=1,m=0$ TE multipole is concentrated in a torus-like shape. Currents proportional to a multipole field near the origin generate the multipole in space so the multipole fields are prescriptions for such current sources. A spherical  layer can couple to a multipole source centered at the origin and enhance its radiation. For example, an oscillating point dipole or a molecule, which emits a TM $l=1$ multipole, can couple to a spherical dielectric layer, similarly to a molecule in an optical cavity \cite{chikkaraddy2016single,thompson2013coupling,gersten1991radiative}. An $l=1$ TE multipole near the origin also generates the same multipole field which can excite a spherical layer that will drive the multipole currents near the origin. This analysis is valid also for the coupling between $l>1$ TE and TM multipoles and a spherical layer. Finally, if a spherical layer or a sphere is excited with a given multipole it can be coupled to another concentric polarizable spherical layer or sphere.

\bibliography{bib}   
\bibliographystyle{apsrev}   
\end{document}